# Implementation and Performance Evaluation of CMOS-integrated Memristor-driven Flip-flop Circuits

Paras Tiwari, Narendra Singh Dhakad, Mohit Kumar Gautam, Shalu Rani, Sanjay Kumar, and Themis Prodromakis

***Abstract*: In this work, we report the implementation and performance evaluation of memristor-driven fundamental logic gates, including NOT, AND, NAND, OR, NOR, and XOR, and novel and optimized designs of the sequential logic circuits, such as D flip-flop, T-flip-flop, JK-flip-flop, and SR-flip-flop. The design, implementation, and optimization of these logic circuits were performed in SPECTRE in Cadence Virtuoso and integrated with 90 nm CMOS technology node. Additionally, we discuss an optimized design of memristor-driven logic gates and sequential logic circuits, and draw a comparative analysis with the other reported state-of-the-art work on sequential circuits. Moreover, the utilized memristor framework was experimentally pre-validated with the experimental data of $Y_2O_3$-based memristive devices, which shows significantly low values of variability during switching in both device-to-device (D2D) and cycle-to-cycle (C2C) operation. The performance metrics were calculated in terms of area, power, and delay of these sequential circuits and were found to be reduced by more than ~24%, 60%, and 58%, respectively, as compared to the other state-of-the-art work on sequential circuits. Therefore, the implemented memristor-based design significantly improves the performance of various logic designs, which makes it more area and power-efficient and shows the potential of memristors in designing various low-power, low-cost, ultrafast, and compact circuits.**

## 1. INTRODUCTION

The rapid global adoption of artificial intelligence (AI) and machine learning (ML) has placed unprecedented demands on modern computing hardware. Conventional complementary-metal-oxide-semiconductor (CMOS)-based digital systems, despite decades of scaling and optimization, face fundamental limitations when applied to AI workloads that require massive parallelism, high memory bandwidth, and energy-efficient data movement [1].

The von Neumann bottleneck, stemming from the physical separation of computation and memory, results in significant overhead due to constant data shuttling between memory and processing units [1-2]. As AI models grow in size and complexity, this bottleneck increasingly dominates power consumption and latency, creating a critical need for alternative hardware paradigms capable of supporting compute-in-memory and low-energy operations [3].

Memristors have emerged as one of the most promising post-CMOS nanodevices for overcoming these bottlenecks [4]. As the fourth fundamental passive circuit element, a memristor exhibits a unique relationship between charge and magnetic flux, giving it the ability to “remember” its resistance state even in the absence of applied power, thereby demonstrating its non-volatile property. The memristor intrinsically combines nonvolatile memory and analog/digital computation within the same nanoscale device. This nonvolatile resistive switching originates from ionic drift, filament formation, or oxygen vacancy migration in nanoscale materials, phenomena that scale favourably with device dimensions. Unlike CMOS transistors, whose leakage currents increase significantly with scaling, memristors exhibit ultra-low leakage and are capable of multi-state storage. Their nanoscale footprint enables extremely high device densities, and their compatibility with back-end-of-line (BEOL) fabrication allows stacking above CMOS layers, leading to true 3D integration. Its ability to store state as resistance, retain information without continuous power, and perform logic operations directly through resistive switching presents a powerful departure from CMOS-only logic implementations.

Despite extensive work on memristor-based combinational logic, including NAND, NOR, XOR, and implication-style logic families, relatively limited research has been conducted on its application to sequential logic elements, which are fundamental for registers, counters, memory buffers, and finite-state machines used in AI datapaths and control units [5-9]. Demonstrating reliable memristor-driven flip-flops and storage elements is therefore crucial for enabling full-stack memristive digital systems. The non-volatile nature of memristor-driven flip-flops could enable instant-on digital systems, reduce checkpointing overhead, and significantly improve power-down power-up resilience for embedded and edge-AI platforms.

Early studies established the feasibility of integrating logic and storage using stacked memristor-CMOS structures, enabling state-retentive logic primitives that form the foundation of non-volatile sequential circuits [7-13].

P. Tiwari and S. Rani are with the Department of Electronics Engineering, Indian Institute of Technology (ISM) Dhanbad, Jharkhand, India

N. S. Dhakad is with Intel Technology India Pvt. Ltd., Bangalore 560103, India.

M. K Gautam is with the School of Electronics and Computer Science, University of Southampton, Southampton, UK.

S. Kumar is with the Department of Electrical Engineering, Indian Institute of Technology Patna, Bihta, Patna 801106, India.

S. Kumar and T. Prodromakis are with the Centre for Electronics Frontiers, School of Engineering, The University of Edinburgh, Edinburgh, Scotland, UK.

P. Tiwari and N. S. Dhakad contributed equally.

*Corresponding authors: sanjaysihag91@gmail.com (S. Kumar); shalu@iitism.ac.in (S. Rani); and t.prodromakis@ed.ac.uk (T. Prodromakis)

Subsequent works demonstrated that memristor-inspired and memristor-based combinational logic circuits can achieve substantial reductions in power and area compared to CMOS technologies, while also addressing variability concerns through optimized device characteristics [11-12]. Building on these advances, several researchers proposed non-volatile latch designs, including optimized pure-memristor latches and hybrid memristor–CMOS D latches, which exhibit lower static power, reduced device count, and improved energy efficiency, making them suitable for low-power sequential applications [9-11]. More recent designs further enhanced latch performance by focusing on area efficiency, energy–delay trade-offs, and detailed comparative analysis with CMOS counterparts [10]. At the flip-flop level, innovative architectures such as D and JK flip-flops demonstrated compact implementation and flexible timing behavior [7-13]. Additionally, ReRAM-based non-volatile latches have been shown to provide strong radiation hardness, extending the applicability of memristive sequential elements to space and safety-critical environments [12]. Overall, the literature indicates that memristor-based latches and flip-flops offer significant advantages in power, area, non-volatility, and functionality, although challenges related to variability, endurance, and large-scale integration remain active research topics [4-13].

On the other hand, previously we explored various memristor-inspired combination logics [14-15], including half/full adder, half/full subtractor, decoder, encoder, multiplexer, and various fundamental logic gates by using experimentally pre-validated memristor framework [16], which showed remarkable performance improvements as compared to the existing CMOS technology and other reported memristor-based state-of-the-art work. Therefore, by considering the aforementioned needs, herein, we extend the memristor logic paradigm by designing and analyzing a complete set of both logic gates and sequential digital circuits using a hybrid memristor-CMOS approach. Building upon the memristive device characteristics and logic methodology established in our previous study, we introduce efficient designs for inverter, NAND, NOR, and XOR gates, and also further develop memristor-driven implementations of essential sequential elements, including D, T, JK, and SR flip-flops. Here, it should be noted that the utilized memristor framework was thoroughly experimentally validated with the experimental data of $Y_2O_3$-based memristive devices as reported in our previous work [16]. These circuits are evaluated for functional correctness, switching behavior, and potential advantages in area efficiency and power reduction. By combining non-volatile resistive memory with CMOS signal restoration, the proposed architectures offer a practical pathway toward low-power, high-density digital subsystems suitable for integration into future AI computing platforms. This expanded design space, encompassing both combinational and sequential logic, lays the groundwork for developing full-scale memristor-augmented digital processors and neuromorphic systems. The insights gained here contribute to the broader effort of bridging memory and computation, reducing idle power, and enabling the ultra-efficient hardware required for the continued advancement of AI technologies.

This article is organized as follows. Section 2 introduces the experimentally validated memristor framework and the physical interpretation of the utilized parameters in modelling. Section 3 describes the design methodology and implementation of memristor-driven combinational logic gates and sequential circuits (flip-flop). This section also describes the various performance metrics, such as area, power, and delay, and their comparison with other state-of-the-art work with existing 90 nm CMOS technology. Section 4 summarizes the conclusion and future scope of the research work

## 2. Memristor Framework

Herein, a detailed explanation and parameter interpretation were provided for the utilized memristor framework [16], which was dually pre-validated with the experimental data of $Y_2O_3$ and $WO_3$-based memristive devices [16] (see Fig. S1 in supporting information). Equation (1) describes the current-voltage (I-V) relation of the utilized memristor framework, wherein several constants were utilized. Table I shows the parameter values and physical interpretation of these parameters as used in Equations (1) to (3).

$$I(t) = \begin{cases} b_1 w^{a_1}\left(e^{\alpha_1 V_i(t)} - 1\right) + \chi\left(e^{\gamma V_i(t)} - 1\right), & \mathrm{V_i(t)} \geqslant 0 \\ b_2 w^{a_2}\left(e^{\alpha_2 V_i(t)} - 1\right) + \chi\left(e^{\gamma V_i(t)} - 1\right), & \mathrm{V_i(t)} < 0 \end{cases} \quad (1)$$

The first term on the right side of Equation (1) shows flux-controlled memristive behaviour, and the second term denotes the ideal diode behaviour in I-V characteristics. Herein, $V_i(t)$ is the applied input voltage.

Equation (2) represents the piecewise window function, *f(w),* which ensures that the state variable (*w*) is varied between 0 and 1. Additionally, the range of parameter *p* defines the boundary of the window function (*f(w))* between 0 and 1, and if *p* is greater than 10, the upper value of *f(w)* is more than 1, which does not follow the essential conditions for the window function, as reported by Prodromakis *et al* [17].

$$f(w) = \log\begin{cases} (1+w)^p, & 0 \leq w \leq 0.1 \\ (1.1)^p, & 0.1 < w \leq 0.9 \\ (2-w)^p, & 0.9 < w \leq 1 \end{cases} \quad (2)$$

The time derivative of the state variable, *w(t),* is shown by Equation (3), which also depends on the nature of the input voltage and window function.

$$\frac{d\omega}{dt} = A \times v_i^m(t) \times f(w) \quad (3)$$

Equation (3) consists of two constants, *A* and *m,* and these are the two independent variables that affect the time derivative of the state variable. Therefore, to ensure that the opposite polarity of the applied voltage results in the opposite change in the rate of change of the state variable, parameter *m* must always be an odd number. Additionally, the previously described models [17-18] are associated with bipolar memristive systems, while the utilized memristor framework was well-suited for both unipolar and bipolar memristive systems, which further increases the flexibility of the memristor framework.

TABLE I: Physical interpretation and numerical values of parameters used in modeling.

| Parameters | Numerical Values | Physical Interpretation |
|---|---|---|
| $b_1$ | $1.59\times10^{-3}$ | Experimental fitting parameters |
| $b_2$ | $-6.2\times10^{-4}$ | |
| $a_1$ | 1.2 | Degree of influence of the state variable under positive bias |
| $a_2$ | 0.3 | Degree of influence of the state variable under negative bias |
| $\alpha_1$ | 0.60 | Hysteresis loop area controlling parameters under positive bias |
| $\alpha_2$ | -0.68 | Hysteresis loop area controlling parameters under negative bias |
| $\chi$ | $1\times10^{-11}$ | Magnitude of ideal diode behavior |
| $\gamma$ | 1 | Diode parameters like thermal voltage and ideality factor |
| $A$ | $5\times10^{-4}$ | Control the effect of the window function. |
| $m$ | 5 | Control the effect of input on the state variable |
| $p$ | 2 ($0 < p \leq 10$) | Bounding parameter for the window function between 0 and 1 |

## 3. Circuit Design Methodology and Results and Discussion

### 3.1 Circuit Design Methodology: Logic Gates

#### Inverter

The memristor-based inverter is the fundamental building block for all other gates. The proposed design utilizes a hybrid CMOS-memristor architecture as shown in Fig. 1(a), consisting of a single memristor and one NMOS transistor. For the inverter shown in Fig. 1(a), one terminal of the memristor is connected to $V_{DD}$, when $V_{IN} = 1$ (high input voltage), the NMOS transistor turns "ON", and the memristor is in the "SET" condition (high resistance). While the NMOS is in the saturation mode, wherein the "ON" resistance is nearly equal to 0. Hence, the output becomes 0. On the other hand, when $V_{IN} = 0$ (low input voltage), the transistor switches into the "cutoff" region; however, the memristor holds the previous resistance value (high), and hence, $V_{OUT}$ is nearly equal to "1," followed by some voltage drop across the memristor.

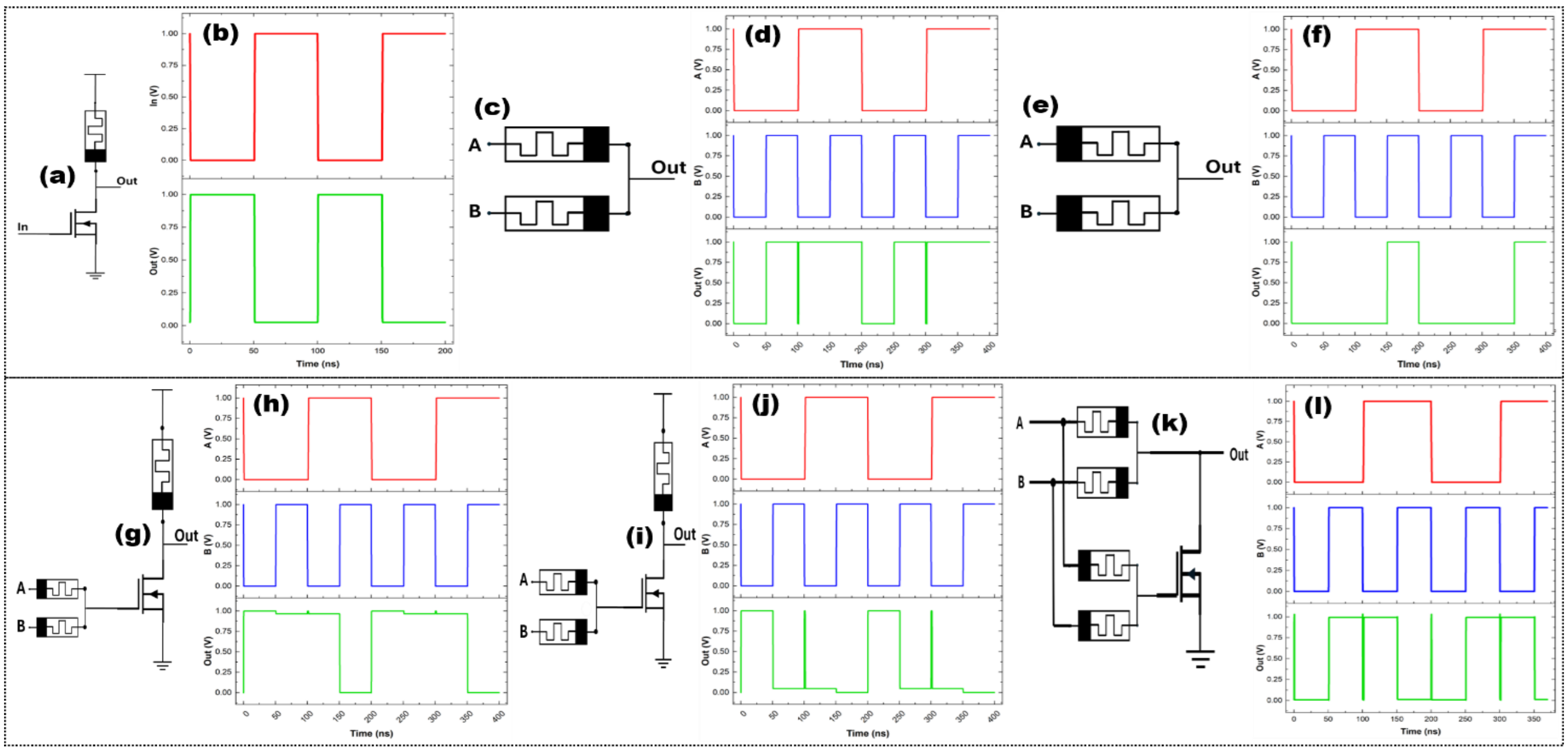

Fig. 1: (a) NOT gate (Inverter), (b) Input and output waveforms of NOT gate, (c) OR gate, (d) Input and output waveforms of OR gate, (e) AND gate, (f) Input and output waveforms of AND gate, (g) NAND gate. (h) Input and output waveforms of NAND gate, (i) NOR gate, (j) Input and output waveforms of NOR gate, (k) XOR gate, and (l) Input and output waveforms of XOR gate.

#### OR/AND Gate

To implement an OR gate, two memristors are connected in parallel as shown in Fig.1(c). Here, one terminal of the memristor is connected to input A and B, while the other terminal of both memristors is connected to the common node output. If both inputs are the same, (0,0) or (1,1), both memristors become identical, and hence, there will be no voltage drop between terminals A and B, so the output follows the input 0 and 1, respectively. For the asymmetric input combination, there will be a current flow from one terminal to another based on the input combination. For A=1 and B=0/A=0 and B=1, the current will flow from A→B/B→A, which impacts the memristor's states. Here, the memristor connected to input 1 offers low resistance, and the other offers high resistance, allowing current to flow from A→B/B→A. The voltage drops across the memristor in the low state, resulting in an output voltage of 1. For symmetric input pairs, the output matches the input, while for asymmetric inputs, the output is 1. Fig. 1(d) demonstrates the simulation waveforms for OR logic. A memristor-based AND gate circuit is shown in Fig. 1(e), in which memristors are inverted, resulting in the memristor states being opposite. For symmetric inputs, the output matches the input. For asymmetric inputs, consider the (1,0) case. Now M1 offers high resistance, and M2 offers low resistance. The voltage drops across M2, resulting in 0 V at the output node. Hence, for asymmetric inputs, the output is 0. This behaviour is clearly demonstrated in the results shown in Fig. 1(f).

### NAND and NOR Gate

The NAND and NOR gates are the universal gates from which all other logic functions can be designed. The NAND and NOR gates are designed using three memristors and one PMOS/NMOS transistor, as compared to the conventional design, where two NMOS and two PMOS transistors are used. The NAND gate is implemented using a memristor-based AND gate followed by a memristor-based inverter, as shown in Fig. 1(g). Inputs A and B each drive a memristor (horizontal devices); their right terminals are shorted to form a common internal node. This internal node controls the gate of the NMOS transistor; the NMOS source is at GND, and the drain goes to the output Y. A vertical memristor from Y to ground provides a conditional pull-down path whose resistance depends on the previous programming/current state, assisting discharge when both inputs activate the network. Similarly, the NOR gate is implemented using an OR gate followed by an inverter, as shown in Fig. 1(i). Fig. 1(h) and 1(j) show the simulation waveform for the NAND and NOR gates, respectively.

### XOR Gate

The XOR gate can be implemented using the combination of AND and OR logic as $A \oplus B = (A + B).\overline{(A.B)}$. The circuit implementation of the same is shown in Fig. 1(k). For the input combinations 00, 01, 10, the AND gate produces output 0, which turns “OFF” the NMOS transistor; hence, the output node follows OR logic. While for input 11, the AND gate output becomes 1, which turns “ON” the NMOS transistor, and the output node gets connected to ground, producing output 0. The proposed design reduces the number of components as compared to the traditional CMOS design, which requires 8-12 transistors. The simulation waveform for the XOR gate is shown in Fig. 1(l). The same XOR logic is used in the next section to implement the flip-flops.

## 3.2 Circuits Design Methodology: Sequential Logic Circuits

In this section, D, T, JK, and SR flip-flops are designed using a hybrid memristor-CMOS approach to exploit the advantages of both emerging and conventional technologies. The proposed designs integrate memristors with CMOS logic to achieve reduced transistor count, lower power consumption, and improved area efficiency compared to conventional CMOS-only implementations. Memristors are utilized for their nonvolatile memory characteristics and compact structure, enabling efficient logic realization, while CMOS transistors provide signal restoration and reliable switching performance. The hybrid architecture ensures correct logical functionality of all flip-flop configurations while maintaining compatibility with standard CMOS fabrication processes. Simulation results demonstrate that the proposed hybrid memristor-CMOS flip-flops exhibit improved power–delay characteristics, making them suitable for low-power and high-density sequential circuit applications.

### D Flip-flop

The operation of the proposed D flip-flop ($Q_{n+1} = D$) is based on a controller (or master)-responder configuration formed by cascading two D latches as shown in Fig. 2(a). The first latch (controller latch) is negative-level triggered, while the second latch (responder latch) is positive-level triggered. Memristors are employed only in the realization of inverter stages within the latches, whereas the remaining circuitry is implemented using a CMOS transistor negative-level-triggered latch. While the second latch (responder latch) is positive-level-triggered, the controller latch is active and samples the input data D, while the responder latch remains isolated, holding its previous state. As the clock transitions from low to high, the controller latch is disabled, thereby locking the sampled data, and simultaneously, the responder latch becomes active, transferring the stored value to the output.

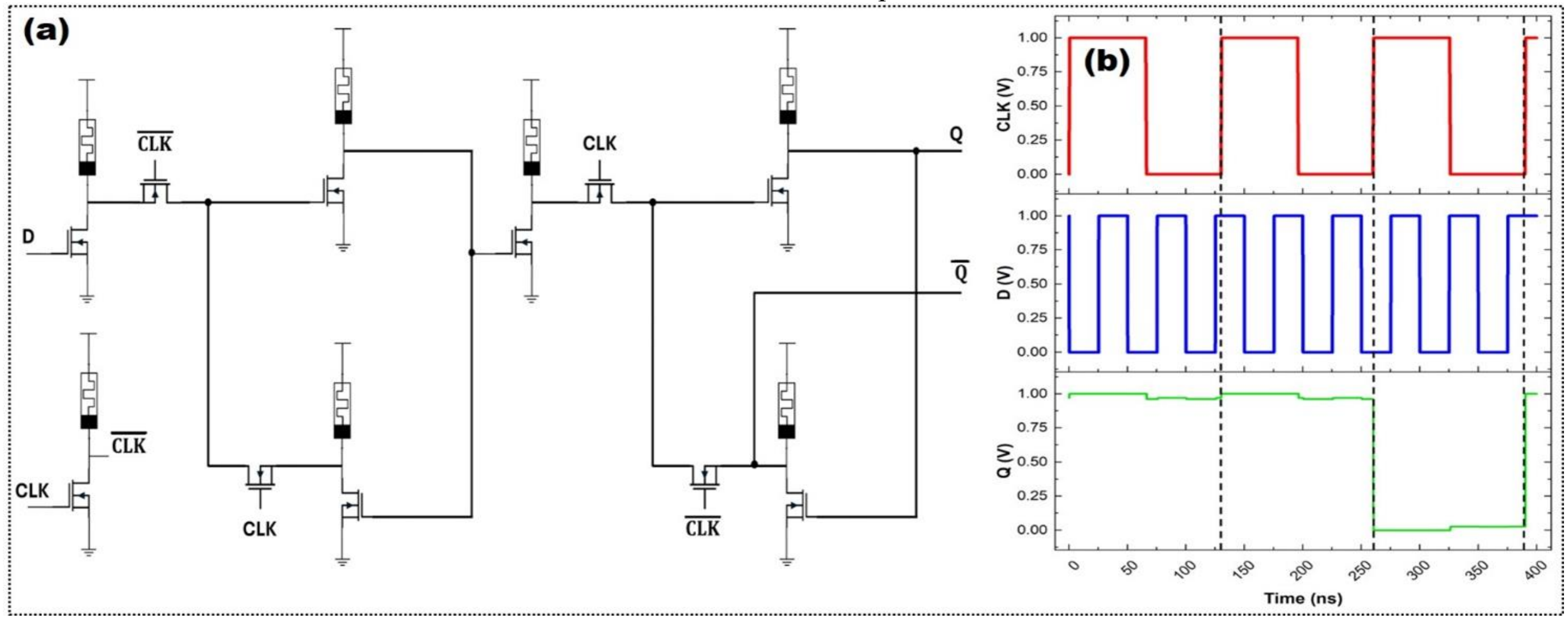

Fig. 2: (a) Circuit schematic of D flip-flop, and (b) CLK, input, and output waveforms of D flip-flop.

During the high level of the clock, changes at the input do not affect the output due to the inactive controller latch. Consequently, the overall configuration exhibits positive edge-triggered behaviour, with the output updating only at the rising edge of the clock signal, ensuring stable and reliable data storage.

## T Flip-Flop

The T flip-flop can be realized using the D flip-flop by incorporating additional combinational logic at the input. For the implementation of the T flip-flop shown in Fig. 3, the D input of the flip-flop is driven by the XOR function of the T input and the present output Q, such that $D = T \oplus Q = T\overline{Q} + \overline{T}Q$. When T input is low, the XOR output remains equal to the current state of Q, causing the D flip-flop to retain its previous state at the next active clock edge. Conversely, when the T input is high, the XOR output becomes the $\overline{Q}$, forcing the D flip-flop to toggle its state on the rising edge of the clock. Since the underlying D flip-flop operates in a positive-edge-triggered manner, the T flip-flop also updates its output only at the clock's rising edge. This approach ensures correct toggle functionality while benefiting from the low-power and area-efficient characteristics of the hybrid memristor-CMOS D flip-flop design.

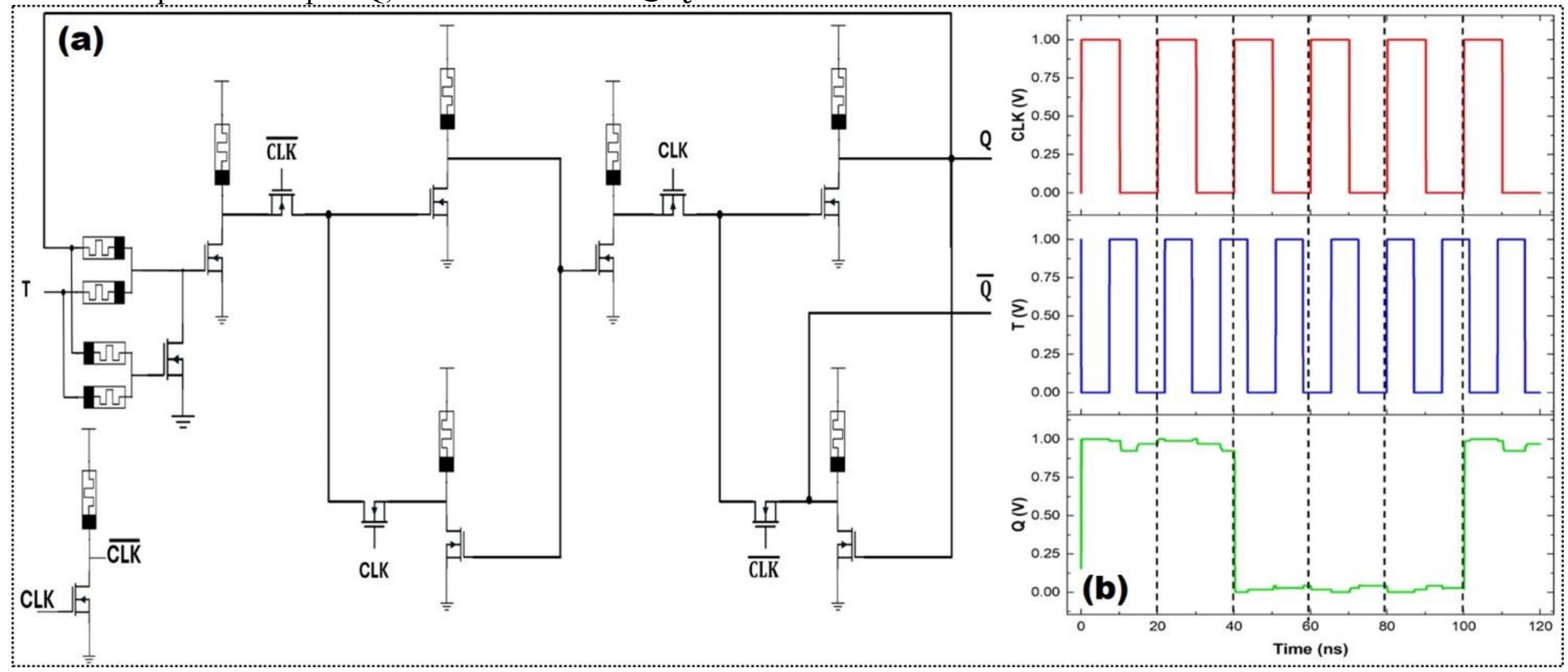

Fig. 3: (a) Circuit schematic of the T flip-flop, and (b) CLK, input, and output waveforms of the T flip-flop.

## SR Flip-Flop

The SR flip-flop is implemented using the proposed $D$ flip-flop by appropriately defining the logic at the $D$ input to realize set and reset operations. In the design shown in Fig. 4, the D input is expressed as $D = S + \overline{R}Q$ where $S$ and $R$ denote the set and reset inputs, respectively, and $Q$ represents the present state of the flip-flop. When the set input $S$ is asserted, the $D$ input is forced high, and the flip-flop sets its output $Q = 1\ (set)$ at the rising edge of the clock. When the reset input $R$ is asserted, the term $\overline{R}$ becomes low, forcing the $D = 0$ and resetting the output $Q = 0\ (reset)$ at the next clock edge. If both $S = R = 0$, the $D$ input follows the present state $Q\ (hold)$, thereby holding the previous output. The underlying positive-edge-triggered $D$ flip-flop ensures that all state transitions occur synchronously with the clock. The invalid condition of $S = R = 1$ is avoided in this implementation to prevent undefined behaviour, resulting in a reliable and clocked SR flip-flop suitable for sequential circuit applications.

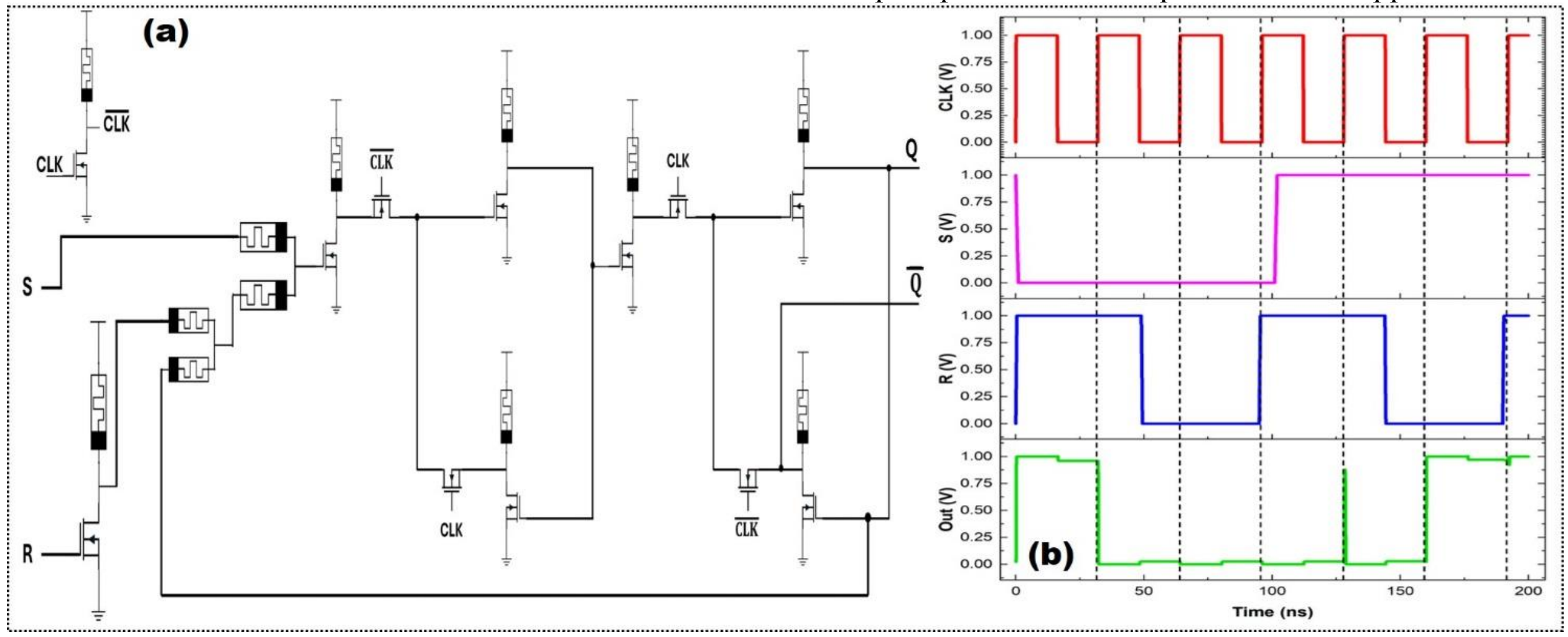

Fig. 4: (a) Circuit schematic of SR flip-flop, and (b) CLK, input, and output waveforms of SR flip-flop.

## JK Flip-Flop

The JK flip-flop is implemented using the proposed D flip-flop by realizing appropriate combinational logic at the $D$ input to achieve the required switching behavior as shown in Fig. 5. In this configuration, the D input is defined as $D = J\overline{Q} + KQ$. When $J = K = 0$, the $D$ input follows the current value of $Q$ $(hold)$, and the flip-flop holds its previous state at the next rising clock edge. When $J = 1, K = 0$, the first term dominates, forcing the $D$ input high and setting $Q = 1$ $(set)$ the output. Conversely, when $J = 0, K = 1$, the second term becomes inactive, driving the $D$ input low and resetting $Q = 0$ $(reset)$ the output. When both $J = K = 1$, the $D$ input becomes $\overline{Q}$ $(toggle)$, causing the flip-flop to toggle its state. Since the core $D$ flip-flop operates in a positive edge–triggered manner, all state transitions of the JK flip-flop occur synchronously at the rising edge of the clock, eliminating race conditions and ensuring stable operation within the hybrid memristor–CMOS framework.

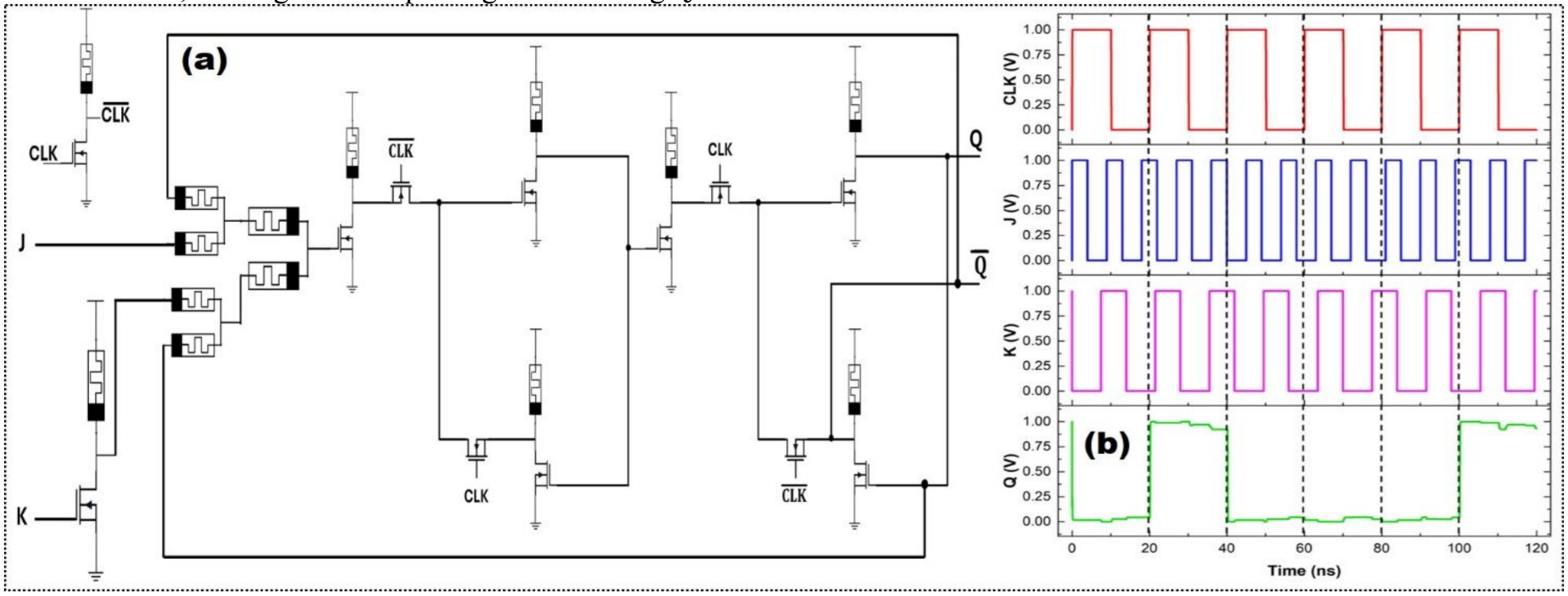

Fig. 5: (a) Circuit schematic of the JK flip-flop, and (b) CLK, input, and output waveforms of the JK flip-flop.

## 3.3 Results and Discussion

In this section, we thoroughly evaluated and discussed the performance of the proposed memristor-based various sequential circuits using SPECTRE and compared them with the previously reported state-of-the-art memristor-based circuit designs as well as CMOS-based circuit designs. Herein, Table II represents a comparative analysis of the proposed D latch with previously implemented and reported data (Ref. 9-12) and a conventional CMOS-based design [12]. The comparison was carried out in terms of total component count used in the specific circuit design, average power consumption, and delay. In terms of transistor usage, the proposed design utilizes 6 transistors**,** which is comparable to existing designs and lower than that reported in [10, 12]. The CMOS design relies entirely on transistors (8 transistors) and does not use memristors, highlighting the advantage of hybrid memristor-CMOS architectures in reducing active device count.

A significant improvement was observed in the average power consumption**.** The proposed design exhibits the lowest power consumption of 7.1 μW**,** which was substantially lower than the other reported state-of-the-art-work [6, 9-12]. Compared to Rasheed e*t al* [9], Rziga *et al* [10], and Yan *et al* [11], the power reduction was approximately 66%, 74%, and 76%**,** respectively. Additionally, as compared with the CMOS-based logic implementation [6], the proposed design demonstrates a power reduction of more than 50%. This significant reduction can be attributed to the use of memristors, which offer non-volatile behavior and reduced switching activity, thereby minimizing dynamic power dissipation. The outcomes clearly indicate that the proposed design was well-suited for low-power and energy-efficient memory and computing applications. The propagation delay of the proposed circuit was measured at 219 ps**,** which was higher than some aggressively optimized designs [6, 9] that report delays in the order of a few picoseconds. However, the majority of previous data have not reported delay values, which limits direct comparison. Moreover, as compared with the CMOS implementation, which exhibits a delay of 10,000 ps**,** the proposed design achieves an improvement of nearly two orders of magnitude**.** Although the delay is not the minimum among all designs, it represents a reasonable trade-off considering the substantial gains in power efficiency and reduced reliance on purely transistor-based logic. Additionally, a first-order technology scaling approach was adopted, where delay was assumed to scale linearly with feature size, while power was assumed to scale with the 1.5$^{th}$ power of the feature size, and detailed analysis was presented in supporting information.

TABLE II: Performance evaluation and comparison of the D latch with other state-of-the-art work.

| Parameters | This work | Ref [9] | Ref [10] | Ref [11] | Ref [12] | Ref [9] [Only CMOS] |
|---|---|---|---|---|---|---|
| Total no. of Components | **10** | 8 | 8 | 9 | 14 | 8 |
| No. of Transistors | **6** | 5 | 7 | 7 | 12 | 8 |
| No. of Memristors | **4** | 3 | 1 | 2 | 2 | Not Required |

| *Avg. Power Consumption (μW) | **7.1** | 71.32 | 78.11 | 29.45 | 126.48 | 51.21 |
|---|---|---|---|---|---|---|
| *Delay (ps) | **219** | 4.50 | Not Reported | Not Reported | 4.46 | 22500 |

*Scaled for 90 nm technology node for fair comparison

Table III represents the performance evaluation of the proposed memristor-based D flip-flop and its comparison with the previously reported designs [7-9, 13]. A significant improvement was observed in average power consumption**,** as compared with other reported data, and the proposed design consumed only 14.2 μW. As observed, the proposed D flip-flop design was more power-efficient than designs reported by Dai *et al* [7], Rasheed e*t al* [9], and Cho *et al* [13], and showed power reduction of more than 60% [7, 9] and about 39% [13], respectively. Additionally, the propagation delay of the proposed D flip-flop was 209.5 ps, showing a clear improvement as compared to other reported literature [4]. However, in our case, the propagation delay was slightly higher than reported by Cho *et al* [13]. Overall, the outcomes demonstrate that the proposed design achieves an effective trade-off between low power consumption and delay.

TABLE III: Performance evaluation and comparison of the D flip-flop with other state-of-the-art work.

| **Parameters** | **This work** | **Ref. [7]** | **Ref. [8]** | **Ref. [9]** | **Ref. [13]** |
|---|---|---|---|---|---|
| Total no. of Components | **18** | 25 | 6 | 14 | 18 |
| No. of Transistors | **11** | 12 | 1 | 9 | 10 |
| No. of Memristors | **7** | 13 | 5 | 5 | 8 |
| *Avg. Power Consumption (μW) | **14.2** | 101.12 | 19.67 | 118.64 | 23.4 |
| *Delay (ps) | **209.5** | 914.4 | Not Reported | Not Reported | 144.6 |

*Scaled for 90 nm technology node for fair comparison

Similarly, Table IV represents the performance evaluation of the proposed memristor-based JK flip-flop and its comparison with previous state-of-the-art work [8, 13]. It should be noted that the proposed JK flip-flop achieves a significant reduction in average power consumption**,** dissipating only 14.2 μW**,** which was substantially lower than that reported by Wang *et al* [8] and Cho *et al* [13], and it was overall improved by approximately 74% [11] and 90% [13], respectively. This highlights the effectiveness of memristor integration in minimizing power dissipation. Moreover, the propagation delay of the proposed design was 147 ps**,** which was significantly lower than previously reported data [13]. On the other hand, we thoroughly examined all the performance parameters for both T and SR, as discussed in Table V, but we were unable to compare their performance with any existing memristor-based circuits due to the non-availability of suitable literature. Therefore, the presented work can add significant advantages to the literature and pave a new path in designing fast and power-efficient sequential circuits for memory and computation applications. Furthermore, Fig. 6 shows the average delay, power, and power-delay-product (PDP) for all the proposed memristor-based sequential circuits. Additionally, the utilized delay formulation is given in supporting information (see Fig. S2).

TABLE IV: Performance evaluation and comparison of the JK flip-flop with other state-of-the-art work.

| **Parameters** | **This work** | **Ref. [8]** | **Ref. [13]** |
|---|---|---|---|
| Total no. of Components | 26 | 9 | 34 |
| No. of Transistors | 12 | 2 | 16 |
| No. of Memristors | 14 | 7 | 18 |
| *Avg. Power Consumption (μW) | 14.2 | 54.59 | 140.3 |
| *Delay (ps) | 147 | Not Reported | 206.8 |

*Scaled for 90 nm technology node for fair comparison

TABLE V: Performance evaluation parameters of proposed memristor-based sequential circuits in this work.

| **Parameters** | **Sequential circuits configuration [This work]** | | | | |
|---|---|---|---|---|---|
| | **SR FF** | **D FF** | **JK FF** | **T FF** | **D latch** |
| Total no. of Components | 24 | 18 | 26 | 23 | 10 |
| No. of Transistors | 12 | 11 | 12 | 12 | 6 |
| No. of Memristors | 12 | 7 | 14 | 11 | 4 |
| *Avg. Power Consumption (μW) | 33.8 | 14.2 | 14.2 | 40.74 | 7.1 |
| *Delay (ps) | 239.5 | 209.5 | 147 | 230 | 219 |

*Scaled for 90 nm technology node for fair comparison

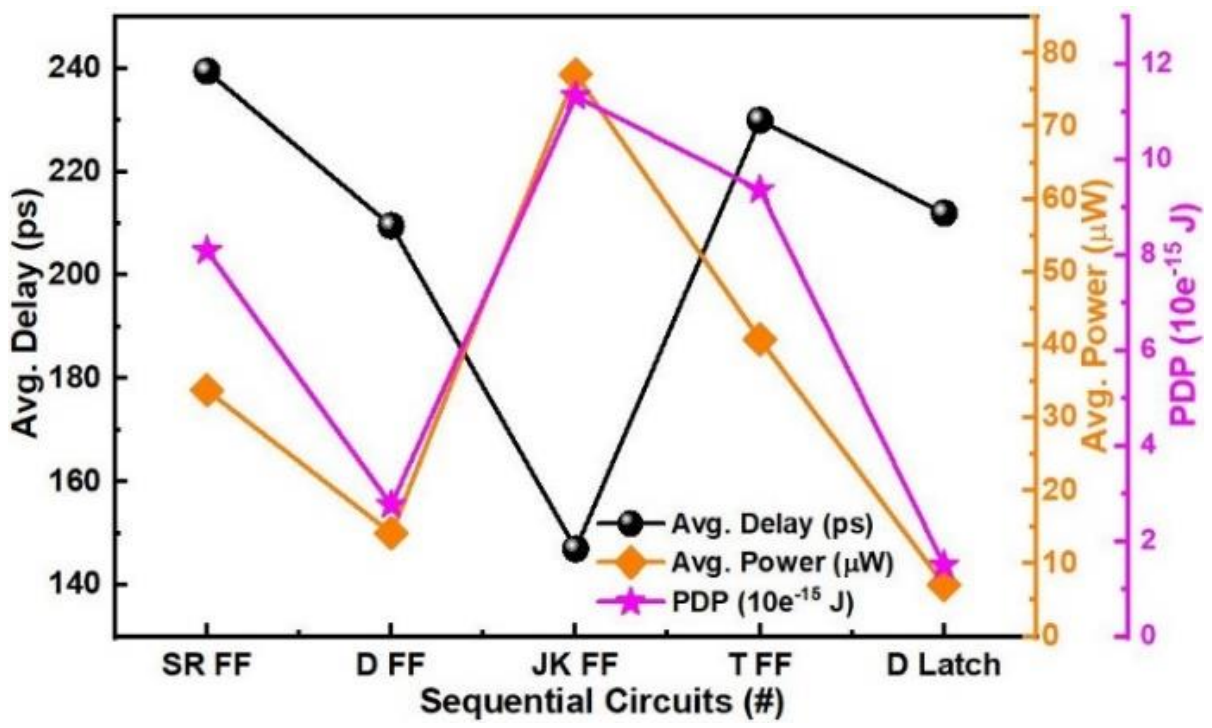

Fig. 6: Delay, power, and PDP analysis of the various proposed combinational circuits.

Fig. S3-S5 in supporting information show the transient response analysis of the logic circuits which were used to design and implement various FF circuits. These analyses include corner analysis (see Fig. S3), temperature variation (see Fig. S4), and voltage variation (see Fig. S5). Fig. S6 shows the voltage transfer characteristics of a single unity-gain transition along with a detailed discussion on the set-up time, hold time, and noise margin.

Additionally, energy-delay product (EDP) was calculated for the designed circuits and presented in Table S1 in supporting information.

## 4. Conclusion

In summary, we investigated CMOS-integrated memristor-driven logic gates and sequential logic circuits (flip-flop), which were designed and implemented in the industry-standard Cadence Virtuoso. The utilized memristor framework was experimentally pre-validated and showed low variance, as confirmed by in-house fabricated memristive devices. Moreover, the proposed memristor-driven logic circuits were integrated with 90 nm CMOS technology, which showed significant improvement in performance in terms of the usage of the number of components, total circuit area, and utilized power as compared to the other state-of-the-art work reported in the literature. Overall, various performance metrics such as area, power, and delay of these sequential circuits were found to be reduced by more than ~24%, 60%, and 58%, respectively. Therefore, the designed sequential circuits are highly reliable for their use in future complex circuits and integrated circuits (ICs). Additionally, this work further opens the possibility to design and implement power, area, and delay-efficient circuits for the registers, counters, amplifier, oscillator, and neural networks.

## Acknowledgements

P. T. would like to thank the IIT (ISM) Dhanbad for providing an institute teaching assistant fellowship for a master's degree. S. K. would like to thank the Department of Science & Technology (DST), New Delhi, for the research grant via IFA23-ENG-375 and Anusandhan National Research Foundation for providing the Prime Minister Early Career Research Grant with grant no. ANRF/ECRG/2025/000636/ENS. S. R. would like to thank the Faculty Research Scheme (FRS) project no. MISC0085 at IIT (ISM) Dhanbad. T. P. would like to thank the EPSRC FORTE Programme (Grant No. EP/R024642/2) and the RAEng Chair in Emerging Technologies (Grant No. CiET1819/2/93) for providing financial support.

## Authors Declaration

Narendra Singh Dhakad is with Intel Technology India Pvt. Ltd., Bangalore, India. This work is undertaken by the author in his personal capacity before joining Intel and is not related to any technical activity at Intel. Any results, conclusions, and opinions presented by the authors are their own and do not in any way represent Intel.

# Supporting Information

# Implementation and Performance Evaluation of CMOS-integrated Memristor-driven Flip-flop Circuits

Paras Tiwari[1], Narendra Singh Dhakad[2], Mohit Kumar Gautam[3], Shalu Rani[1,*], Sanjay Kumar[4,5,*], and Themis Prodromakis[5,*]

[1]Department of Electronics Engineering, Indian Institute of Technology (ISM) Dhanbad, Jharkhand, India
[2]Intel Technology India Pvt. Ltd., Bangalore 560103, India.
[3]School of Electronics and Computer Science, University of Southampton, Southampton, UK.
[4]Department of Electrical Engineering, Indian Institute of Technology Patna, Bihta, Patna 801106, India.
[5]Centre for Electronics Frontiers, School of Engineering, The University of Edinburgh, Edinburgh, Scotland, UK.

P. Tiwari and N. S. Dhakad contributed equally.

[*]Corresponding authors: sanjaysihag91@gmail.com (S. Kumar); shalu@iitism.ac.in (S. Rani); and t.prodromakis@ed.ac.uk (T. Prodromakis)

The utilized memristor framework was dually pre-validated with experimental data of both $Y_2O_3$ (for high switching voltage up to ±5 V) and $WO_3$ (for lower switching voltage up to ±1.5 V), as reported in our previously reported work [1]. As previously discussed, the slight variations in the utilized parameter range were also considered during the experimental validation of the utilized memristor model to handle the device current range from milliampere (mA) to microampere (μA). Here, Fig. S1 depicts both experimental validation for better clarity and understanding. As observed in Fig. R2, the switching threshold of the memristive model is higher than 1.2 V, which allows us to read the memristive state beyond this voltage, and the read disturbance problem is not present.

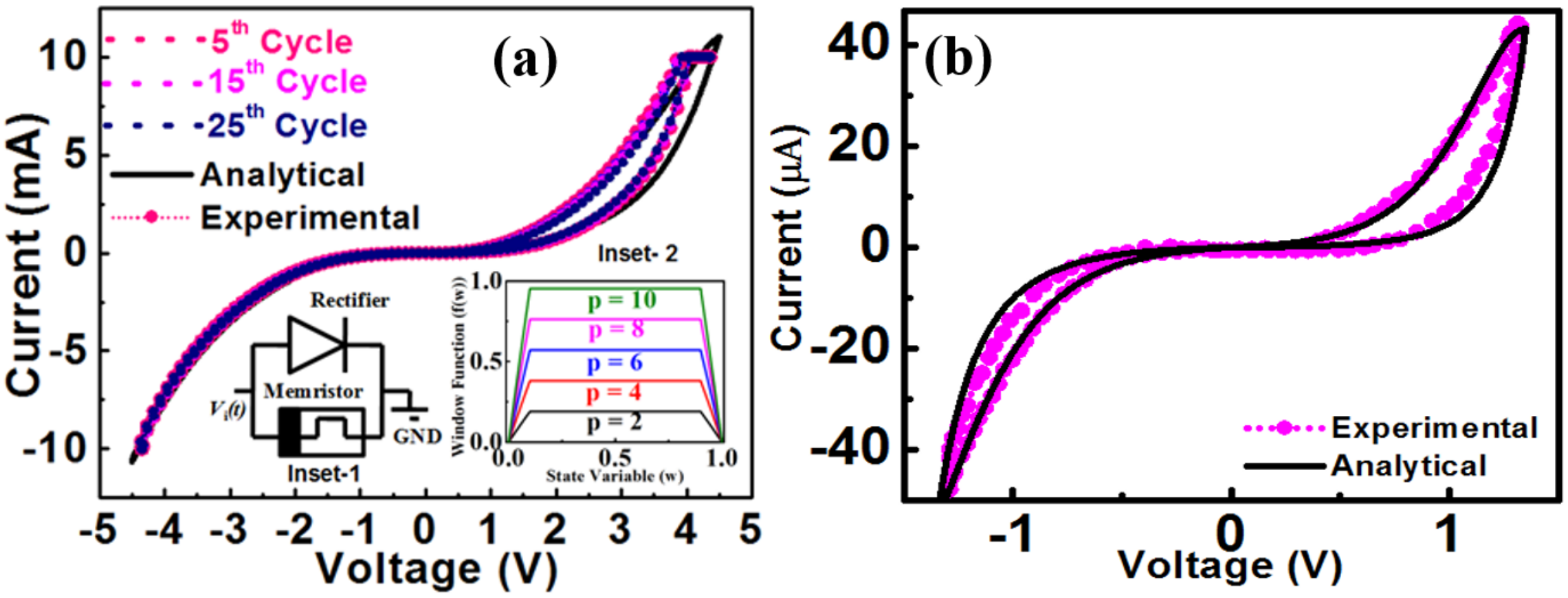

Fig. S1: (a) Resistive switching characteristic of a memristive device (sweep rate: 1.042 V/s) validation with $Y_2O_3$-based memristor. Insets 1 and inset-2 show the parallel connection of the rectifier and memristor and a piecewise window function, respectively, and (b) Resistive switching characteristics of $WO_3$-based memristive system (sweep rate: 2 V/s) [1].

The following delay formulation has been considered for evaluation purposes [2].

The reported average delay = ($T_{phl}$ + $T_{plh}$)/2.

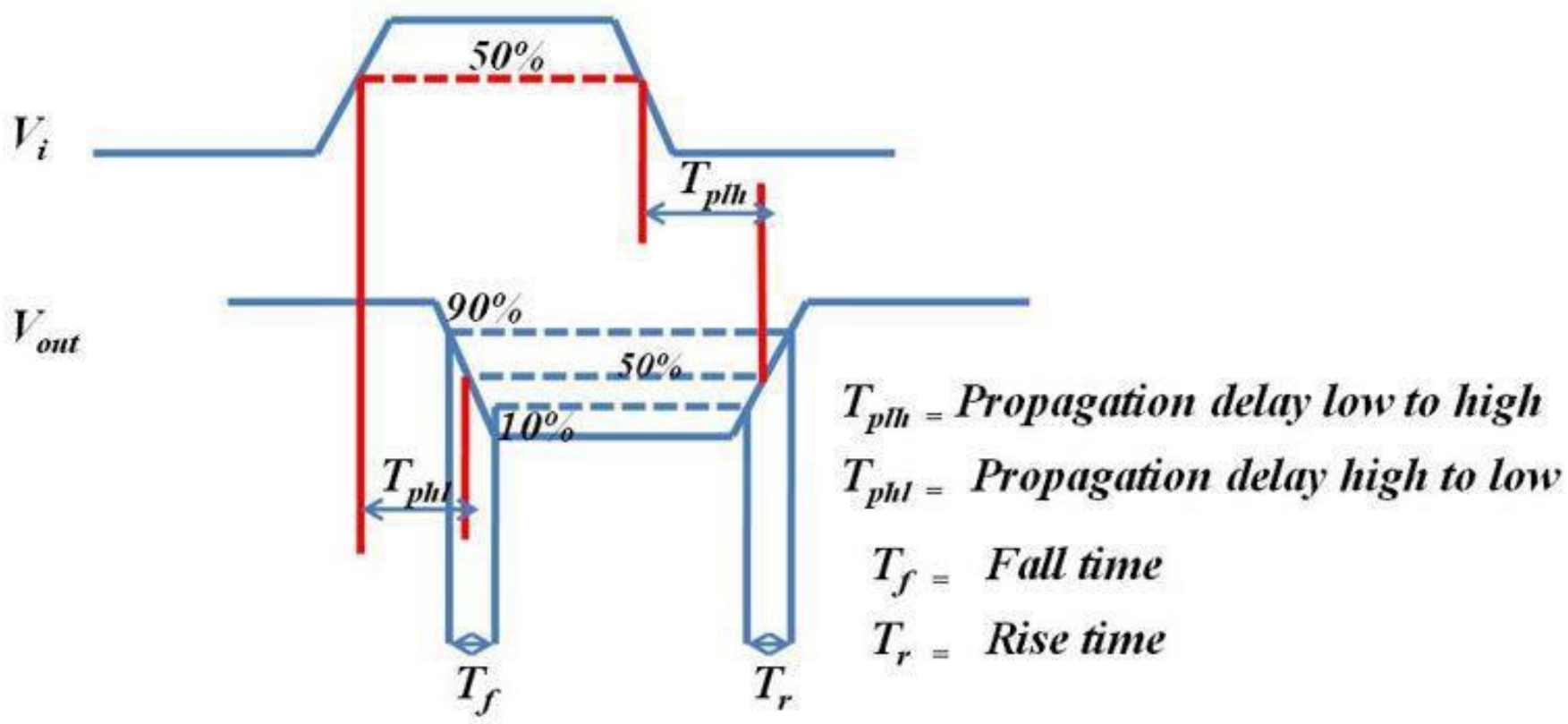

Fig. S2: Schematic of propagation delay calculation.

Further, to enable a fair comparison among designs implemented in different CMOS technology nodes, all reported power and delay values were normalized to an equivalent 90 nm technology node. A first-order technology scaling approach was adopted, where delay was assumed to scale linearly with feature size, while power was assumed to scale with the 1.5th power of the feature size.

The normalized delay and power were calculated using:

$$D_{90} = D_l \times (90/L) \quad (1)$$

$$P_{90} = P_l \times (90/L)^{1.5} \quad (2)$$

where:

$D_l$ and $P_l$ are the delay and power reported at the original technology node $L$ (nm),

$D_{90}$ and $P_{90}$ are the corresponding delay and power values normalized to the 90 nm technology node.

The transient response of the proposed hybrid logic circuits was performed, and using these logic circuits, presented flip-flops were designed and analyzed across different process corners, supply voltages, and temperature conditions to evaluate their robustness and reliability, as shown in Fig. S3-S5. Simulations were performed under typical-typical (TT), slow-slow (SS), fast-fast (FF), slow-fast (SF), and fast-slow (FS) process corners, while varying the supply voltage and operating temperature over the specified ranges. The obtained waveforms demonstrate correct data capture and output transitions in all operating conditions, indicating stable functionality of the flip-flop.

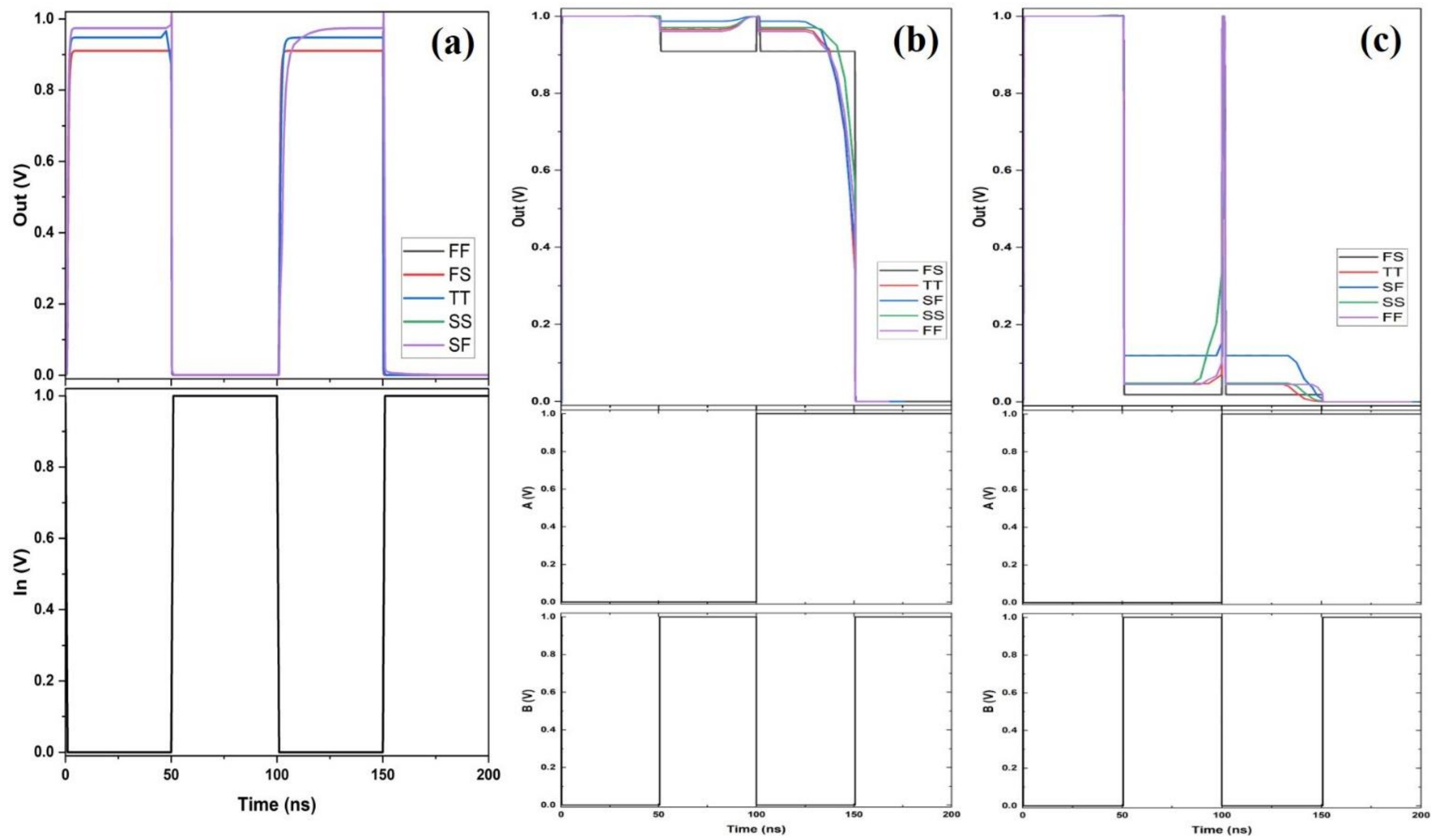

Fig S3: (a) Inverter, (b) NAND and (c) NOR gate transient response for corner analysis.

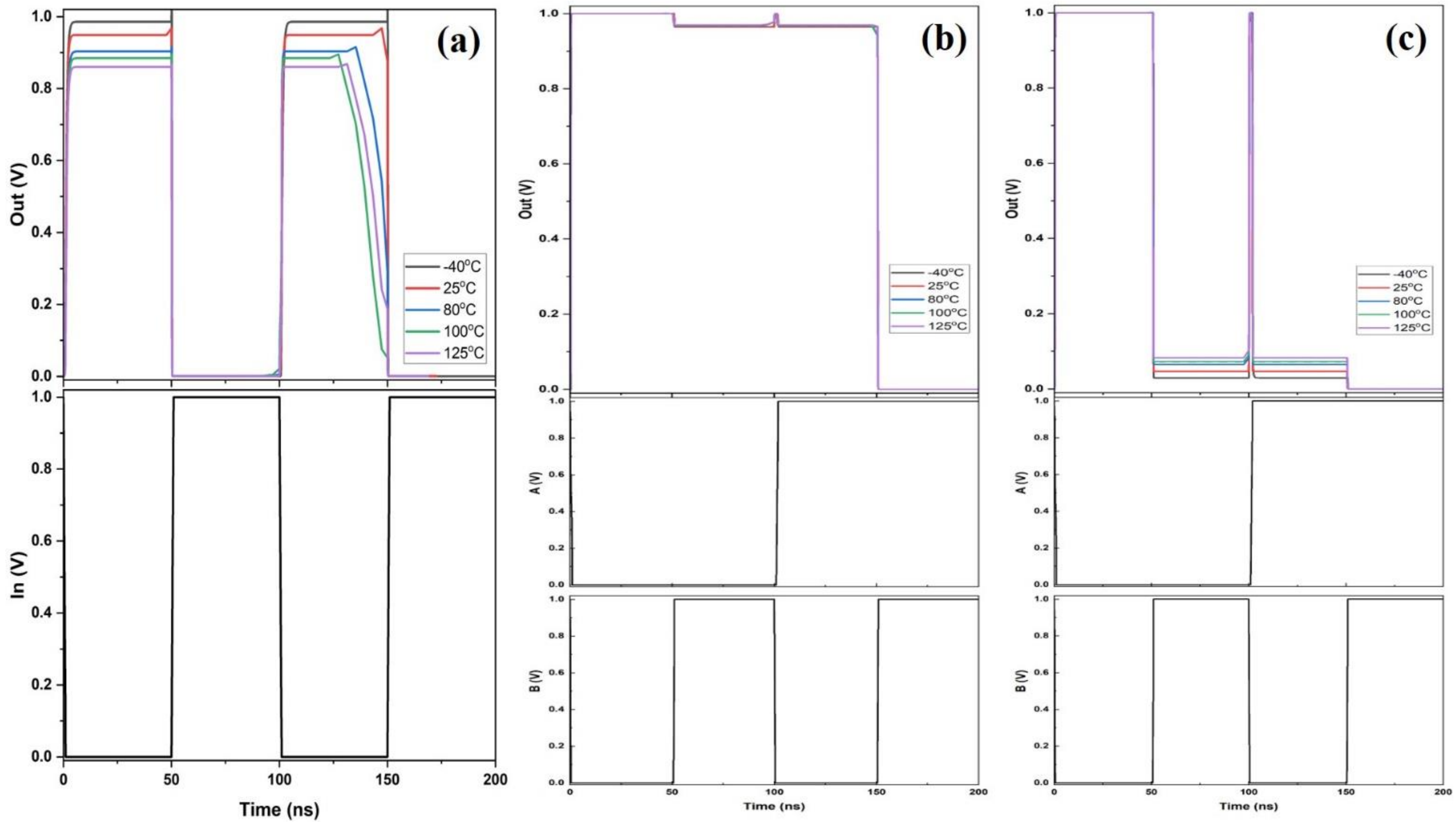

Fig S4: (a) Inverter, (b) NAND and (c) NOR gate transient response for temperature variation.

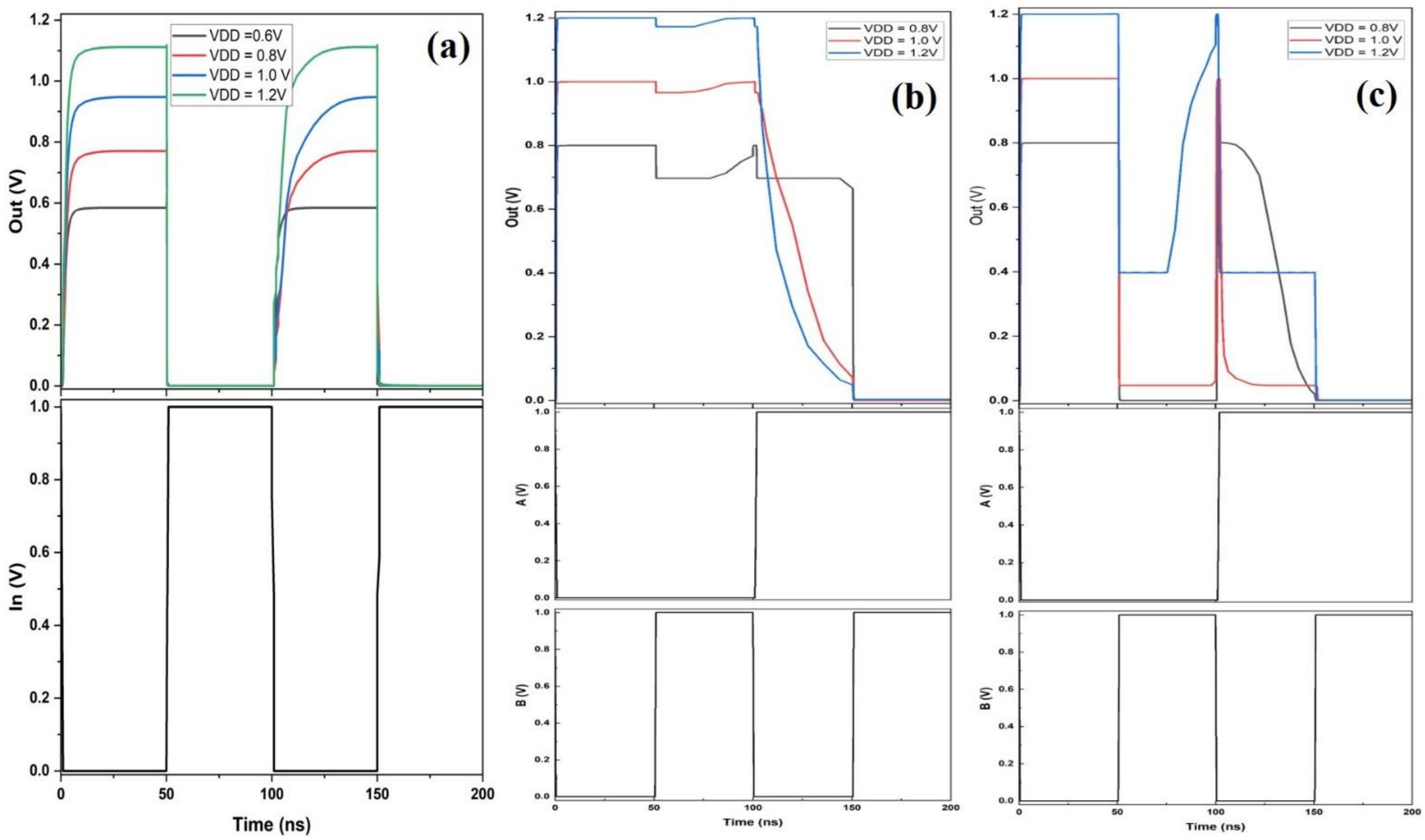

Fig S5: (a) Inverter, (b) NAND and (c) NOR gate transient response for voltage variation.

Energy-delay product (EDP) was calculated for our designed circuits as given in Table S1. Additionally, for the experimental perspective, the expected average energy consumption by the oxide-based memristive devices during ON and OFF conditions is $E_{\mathrm{SET}}$ = 744 pJ and $E_{\mathrm{RESET}}$ = 17.73 fJ, respectively [3].

Table S1: Energy-delay product (EDP) of the various designed FF circuits.

| Flip-flop | Average Power (μW) | Time Period (ns) | Average Energy (fJ) | Delay (ps) | Energy-Delay Product (EDP) $\times 10^{-30}$ J |
|---|---|---|---|---|---|
| D FF | 14.2 | 130 | 1846 | 209.5 | 386.737 |
| T FF | 40.74 | 20 | 814.8 | 230 | 187.404 |
| SR FF | 33.8 | 32 | 1081.6 | 239.5 | 259.0432 |
| JK FF | 14.2 | 20 | 284 | 147 | 41.748 |

Setup and hold times were extracted through data-to-clock timing sweeps under nominal process-voltage-temperature conditions. The setup time was evaluated by sweeping the pre-clock data transition. The hold time was evaluated by sweeping the post-clock data transition. Correct operation was maintained even when the data transition occurred immediately after the clock edge (as the flip-flop is edge-triggered), indicating an approximately hold-time requirement.

The minimum time before the active clock edge that the input $D$ must remain stable for correct sampling.

$$t_{setup} = t_{CLK} - t_{D(last\ transition)} = 1.5\ ns$$

Hold Time: ~0

Clock to Q Time: Delay from the triggering clock edge to the output reaching 50% of its final value = 0.209 ns

Maximum operating frequency = 600 MHz

Due to the resistive switching characteristics of the RRAM device, the voltage transfer characteristic exhibits a single unity-gain transition (see Fig. S6) point rather than the two transition points typically observed in CMOS inverters. Therefore, conventional CMOS noise margin extraction based on $V_{IL}$and $V_{IH}$is not directly applicable.

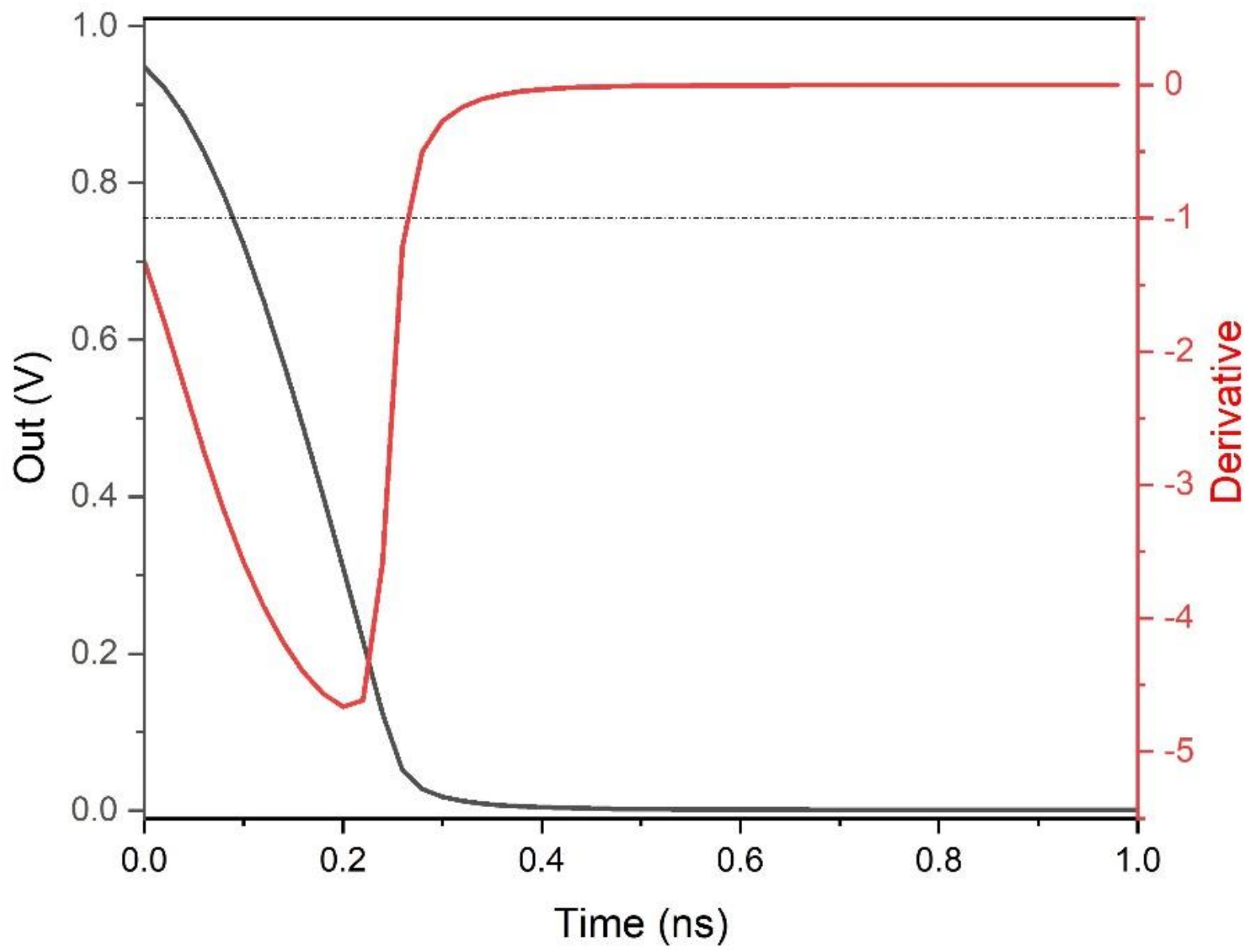

Fig. S6: Voltage transfer characteristic of a single unity-gain transition.

The noise margin of the inverter is calculated by plotting the derivative curve of DC characteristics and calculating the slope at -1 using the equations below.

$$NM_L = V_{IL} - V_{OL}$$
$$NM_H = V_{OH} - V_{IH}$$

Here, for inverter DC characteristics, $V_{IL} - V_{OL} = 0V, V_{OH} = 1\ V, V_{IH} = 0.28.$

Hence $NM_L = 0$, $NM_H$= 0.72 V.